\documentclass[10pt, conference, compsocconf]{IEEEtran}
\IEEEoverridecommandlockouts
\usepackage[utf8]{inputenc}
\usepackage[T1]{fontenc}
\usepackage[english]{babel}
\usepackage{ntheorem}
\usepackage{amsmath,amssymb}
\usepackage{epsfig,psfrag}
\usepackage{graphics,graphicx} 
\usepackage{color}
\usepackage{dsfont}
\usepackage{graphicx}
\usepackage{stmaryrd}
\usepackage{color}
\usepackage[font=footnotesize]{subfig}
\usepackage{algorithm2e}
\renewcommand{\theequation}{\arabic{equation}}
\usepackage[width=18.2cm,height=25.7cm]{geometry}
\usepackage{epsfig}
\usepackage{multirow}

\SetKwInput{KwData}{Parameters}

\begin{document}

\newtheorem{Corol}{Corollary}
\newtheorem{Prop}{Proposition}
\newtheorem{Conjec}{Conjecture}
\newtheorem{lemma}{Lemma}
\newtheorem{Def}{Definition}
\newtheorem{theorem}{Theorem}

\newtheorem{definition}{Definition}
\newtheorem{notation}{Notation}
\newtheorem{proposition}{Proposition}
\newtheorem{corollary}{Corollary}
\newtheorem{remark}{Remark}
\newtheorem{example}{Example}



\newcommand{\rel}[0]{\ensuremath{{\mathcal{R}}}}
\newcommand{\Gall}[0]{\ensuremath{\mathcal{G}}}

\newcommand{\harvey}[0]{\ensuremath{\textsf{haRVey}\xspace}}
\newcommand{\barvey}[0]{\ensuremath{\textsf{barvey}\xspace}}
\newcommand{\bamTorv}[0]{\ensuremath{\textsf{bam2rv}\xspace}}
\newcommand{\rvqe}[0]{\ensuremath{\textsf{rvqe}\xspace}}
\newcommand{\Why}[0]{\ensuremath{\textsf{Why}\xspace}}
\newcommand{\Be}[0]{\textsf{B}\xspace}
\newcommand{\Zed}[0]{\ensuremath{\textsf{Z}\xspace}}

\newcommand{\Lang}[0]{\ensuremath{{\cal L}}}
\newcommand{\Sorts}[0]{\ensuremath{{\cal S}}}
\newcommand{\Vars}[0]{\ensuremath{{\cal V}}}
\newcommand{\Funcs}[0]{\ensuremath{{\cal F}}}
\newcommand{\Preds}[0]{\ensuremath{{\cal P}}}
\newcommand{\arite}[0]{\ensuremath{{\alpha}}} 
\newcommand{\sorte}[0]{\ensuremath{{\sigma}}}
\newcommand{\tauTerm}[0]{\textit{$\tau$-terme}\xspace}
\newcommand{\tauITerm}[1]{\textit{$\tau_{#1}$-term}\xspace}
\newcommand{\atome}[0]{\textit{atome}\xspace}
\newcommand{\fpoms}[0]{\textit{fpoms}\xspace}
\newcommand{\VLibre}[1]{\Vars_{\textit{lib}}(#1)}
\newcommand{\congruenceE}[0]{\ensuremath{{\cal E}}}
\newcommand{\preInterpret}[1]{\ensuremath{[ #1 ]}}
\newcommand{\preInterpretSSET}[1]{\preInterpret{#1}}
\newcommand{\preInterpretBAES}[1]{I'(#1)}
\newcommand{\preinterpret}[0]{\preInterpret{\,}}
\newcommand{\valuationSSET}[1]{\valuation(#1)}
\newcommand{\valuationBAES}[1]{\valuation'(#1)}
\newcommand{\valuationX}[1]{\valuation(#1)}
\newcommand{\valuation}[0]{\ensuremath{\rho}\xspace}
\newcommand{\etat}[0]{\ensuremath{\rho}}
\newcommand{\constt}[1]{\ensuremath{\textsf{#1}}}
\newcommand{\termSet}[0]{t}

\newcommand{\sset}[0]{\ensuremath{\mathit{SSET}}}
\newcommand{\axCod}[0]{\ensuremath{\delta}}
\newcommand{\mty}[0]{\ensuremath{\mathsf{mty}}}
\newcommand{\ttrue}[0]{\ensuremath{\mathsf{tt}}}
\newcommand{\tfalse}[0]{\ensuremath{\mathsf{ff}}}

\newcommand{\interpretI}[0]{\ensuremath{{\cal I}}\xspace}
\newcommand{\classInterpret}[0]{\ensuremath{\mathbf{I}\xspace}}
\newcommand{\ltrue}[0]{\ensuremath{\top}}
\newcommand{\lfalse}[0]{\ensuremath{\bot}}
\newcommand{\modele}[0]{\ensuremath{\Vdash}}
\newcommand{\deduc}[0]{\vdash}

\newcommand{\conseq}[0]{\ensuremath{\vDash}}
\newcommand{\herbrandU}[0]{\ensuremath{\mathbbm{H}}\xspace}

\newcommand{\Nats}[0]{\ensuremath{\mathds{N}}}
\newcommand{\Z}[0]{\ensuremath{\mathbb{Z}}}
\newcommand{\R}[0]{\ensuremath{\mathds{R}}}
\newcommand{\Bool}[0]{\ensuremath{\mathds{B}}}
\newcommand{\StratSet}[0]{\ensuremath{\mathbb{S}}}

\newcommand{\impliq}[0]{\ensuremath{{\Rightarrow}}}
\newcommand{\equival}[0]{\ensuremath{{\Leftrightarrow}}}

\newcommand{\sortedForall}[3]{\ensuremath{(\forall_{#1} #2\, . \, #3)}}
\newcommand{\sortedExists}[3]{\ensuremath{(\exists_{#1} #2\, . \, #3)}}
\newcommand{\sortedQ}[3]{\ensuremath{(Q_{#1} #2\, . \, #3)}}
\newcommand{\Forall}[2]{\ensuremath{(\forall #1\, . \, #2)}}
\newcommand{\qnnf}[0]{\ensuremath{_{\textit{nf}}}}
\newcommand{\gq}{\ensuremath{\mathsf{gq}}}
\newcommand{\gqe}{\ensuremath{\mathsf{gqe}}}
\newcommand{\bdd}[1]{\textit{bdd}(#1)}

\newcommand{\regleSep}[0]{\mathtt{~\textbf{|}~}}
\newcommand{\regleDeriv}[0]{\mathtt{::=}}

\newcommand{\transfo}[0]{\ensuremath{{\rightarrow}}}
\newcommand{\negativeForm}[1]{\ensuremath{\textit{neg}({#1})}}
\newcommand{\cnf}[1]{\ensuremath{\textit{cnf}({#1})}}
\newcommand{\dnf}[1]{\ensuremath{\textit{dnf}({#1})}}
\newcommand{\polar}[1]{\ensuremath{\textit{pol}(#1)}}

\newcommand{\statesVars}[0]{\ensuremath{X}}
\newcommand{\statesSet}[0]{\ensuremath{{\Sigma}}}
\newcommand{\initialAssertion}[0]{\ensuremath{I}}
\newcommand{\transSymbol}[0]{\ensuremath{{\rightarrow}}}
\newcommand{\transLabel}[0]{\ensuremath{{\textit{act}}}}
\newcommand{\prodASync}[0]{\ensuremath{\otimes}}
\newcommand{\prodSync}[0]{\ensuremath{\oplus}}

\newcommand{\execTS}[1]{\ensuremath{[[ #1 ]]}}

\newcommand{\ite}[3]{\textit{ite}(#1,#2,#3)}
\newcommand{\iteRV}[3]{\mathsf{ite}(#1,#2,#3)}
\newcommand{\eqd}[0]{\ensuremath{=_{\textit{\small{def}}}}}
\newcommand{\instruction}[1]{\xspace{\sf #1}\xspace}
\newcommand{\ifB}[0]{\instruction{if~}}
\newcommand{\thenB}[0]{\instruction{~then~}}
\newcommand{\elseB}[0]{\instruction{~else~}}
\newcommand{\skipB}[0]{\instruction{skip}}

\newcommand{\select}[2]{\ensuremath{\mathsf{rd}(#1,#2)}}
\newcommand{\store}[3]{\ensuremath{\mathsf{wr}(#1,#2,#3)}}
\newcommand{\InsO}[0]{\ensuremath{\mathsf{ins}}}
\newcommand{\Ins}[2]{\ensuremath{\InsO(#1,#2)}}
\newcommand{\enumO}[0]{\ensuremath{\mathsf{enum}}}
\newcommand{\enum}[1]{\ensuremath{\enumO(#1)}}

\newcommand{\PRE}[1]{\ensuremath{\langle #1 \rangle}}
\newcommand{\PREB}[1]{\ensuremath{[ #1 ]}}
\newcommand{\Pre}[2]{\ensuremath{\textit{Pre}_{#1}(#2)}}
\newcommand{\pretilde}[0]{\ensuremath{\widetilde{\textit{pre}}}}
\newcommand{\post}[0]{\ensuremath{\textit{post}}}
\newcommand{\pretildeR}[0]{\ensuremath{\widetilde{\textrm{pre}}}}
\newcommand{\postR}[0]{\ensuremath{\textrm{post}}}
\newcommand{\Succ}[0]{\ensuremath{\textit{successeur}}}
\newcommand{\Men}[0]{\ensuremath{\textit{meneur}}}
\newcommand{\Gpre}[2]{\ensuremath{G_{#1}(#2)}}
\newcommand{\Fpre}[2]{\ensuremath{F_{#1}(#2)}}
\newcommand{\POST}[1]{\ensuremath{[ #1 ]^{o}}}
\newcommand{\sem}[1]{\ensuremath{\textit{rel}_{#1}}}
\newcommand{\quantToSubst}[0]{\ensuremath{\textit{q2s}}}

\newcommand{\wpr}[0]{\ensuremath{\textit{wp}}\xspace}
\newcommand{\wlp}[0]{\ensuremath{\textit{wlp}}\xspace}
\newcommand{\weaklp}[2]{\ensuremath{\textit{wlp}_{#1}(#2)}\xspace}
\newcommand{\weakp}[2]{\ensuremath{\textit{wp}_{#1}(#2)}\xspace}
\newcommand{\prd}[2]{\ensuremath{\textit{prd}_{#1}({#2})}\xspace}

\newcommand{\myInput}[1]{\begin{center}\begin{tabular}{l}\input{#1}\end{tabular}\end{center}}

\newcommand{\Init}[0]{\ensuremath{\textit{Init}}\xspace}
\newcommand{\Target}[0]{\ensuremath{\textit{Target}}\xspace}

\newcommand{\substs}[0]{\ensuremath{\textit{Substs}}}
\newcommand{\choice}[0]{\ensuremath{[]}}
\newcommand{\paral}[0]{\ensuremath{\mid\mid}}
\newcommand{\card}[1]{\ensuremath{\textit{card}(#1)}}

\newcommand{\fs}[1]{\ensuremath{\mathsf{#1}}}
\newcommand{\sort}[1]{\mbox{\textsc{#1}}}
\newcommand{\dropE}[0]{\ensuremath{\mathsf{dropExistential}}}
\newcommand{\renameF}[0]{\ensuremath{\mathsf{renameFormula}}}
\newcommand{\ppnf}[0]{\ensuremath{\mathsf{de}}}
\newcommand{\apnx}[0]{\ensuremath{\mathsf{mini}}}
\newcommand{\apnxq}[0]{\ensuremath{\mathsf{m}_q}}
\newcommand{\sqs}[0]{\ensuremath{\mathsf{rf}}}
\newcommand{\theoryT}[0]{\ensuremath{{\cal T}}}
\newcommand{\theoryU}[0]{\ensuremath{{\cal T}}}
\newcommand{\vars}[0]{\ensuremath{\mathcal{V}}}

\newcommand{\latice}[0]{\ensuremath{\textit{exp}^{A}}}
\newcommand{\laticeB}[0]{\ensuremath{\latice(B_1,\ldots,B_l)}}
\newcommand{\concretisation}[0]{\ensuremath{\gamma}}
\newcommand{\abstraction}[0]{\ensuremath{\alpha}}

\newcommand{\sfunc}[0]{\ensuremath{\textit{SFUNC}}}
\newcommand{\override}[2]{\ensuremath{#1 \bdres #2 }}
\newcommand{\Partition}[1]{\ensuremath{\mathbbm{P}}(#1)}
\newcommand{\fc}[1]{\ensuremath{\widetilde{#1}}} 
\newcommand{\nul}[0]{\ensuremath{\mathsf{null}}} 
\newcommand{\overR}[3]{\ensuremath {\mathsf{over}(#1,#2,#3)}}
\newcommand{\image}[2]{\ensuremath {\sf{image}(#1,#2)}}


\newcommand{\ind}[0]{\textit{index}}
\newcommand{\val}[0]{\textit{val}}
\newcommand{\arr}[0]{\textit{array}}
\newcommand{\iSet}[0]{\ensuremath{\textit{S}_{i}}\xspace}
\newcommand{\eSet}[0]{\ensuremath{\textit{S}_{e}}\xspace}
\newcommand{\ARR}[0]{\textit{ARRAY}\xspace}
\newcommand{\IND}[0]{\textit{INDEX}\xspace}
\newcommand{\VAL}[0]{\textit{VALUE}\xspace}
\newcommand{\constArr}{\ensuremath{\mathsf{const}}\xspace}
\newcommand{\im}[0]{\ensuremath{\mathsf{im}}}

\newcommand{\eJ}{\EuScript J}
\newcommand{\eF}{\EuScript F}
\newcommand{\eL}{\EuScript L}
\newcommand{\eT}{\EuScript T}
\newcommand{\eV}{\EuScript V}
\newcommand{\eP}{\EuScript P}
\renewcommand{\le}{\leqslant}
\renewcommand{\ge}{\geqslant}

\newcommand{\any}[0]{\textrm{@}}

\newcommand{\str}[0]{\textit{Strength}}

\newcommand{\sortSet}{\mathcal{S}}
\newcommand{\sortVars}{\mathcal{X}_s}
\newcommand{\sortFunctionSet}{\mathcal{F}_s} 
\newcommand{\varSet}{\mathcal{X}}
\newcommand{\functionSet}{\mathcal{F}} 

\def\unique{\sort{u}}

\author{Jacques M. Bahi, Jean-Fran\c{c}ois Couchot, Christophe Guyeux, and
Qianxue Wang*~\thanks{* Authors in alphabetic order}\\
 \{jacques.bahi, jean-francois.couchot, christophe.guyeux, qianxue.wang\}@univ-fcomte.fr}

\title{Class of Trustworthy Pseudo-Random Number Generators}  
\author{\IEEEauthorblockN{Jacques M. Bahi\IEEEauthorrefmark{1},
Jean-Fran\c{c}ois Couchot\IEEEauthorrefmark{1}, Christophe Guyeux\IEEEauthorrefmark{1}and
Qianxue Wang\IEEEauthorrefmark{1}}
\IEEEauthorblockA{\IEEEauthorrefmark{1}University of Franche-Comte\\
Computer Science Laboratory LIFC,
Belfort, France\\ Email:\{jacques.bahi, jean-francois.couchot, christophe.guyeux, qianxue.wang\}@univ-fcomte.fr}}

\maketitle

\begin{abstract}
With the 
widespread use of communication technologies, cryptosystems are therefore critical to guarantee security over open networks as the Internet.
Pseudo-random number generators (PRNGs) are fundamental in cryptosystems and information hiding schemes.
One of the existing chaos-based PRNGs is using chaotic iterations schemes. 
In prior literature, the iterate function is just the vectorial boolean negation.  
In this paper, we propose a method using Graph with strongly connected components as a selection criterion for chaotic iterate function. 
In order to face the challenge of using the proposed chaotic iterate functions in PRNG, these PRNGs are subjected to a statistical battery of tests, which is the well-known NIST in the area of cryptography.

\end{abstract}

\begin{IEEEkeywords}
Internet security; Chaotic sequences; Statistical tests; Discrete chaotic iterations.
\end{IEEEkeywords}
\IEEEpeerreviewmaketitle

\section{Introduction}\label{sec:intro}
Chaos and its applications in the field of secure communication have attracted a
lot of attention  in various domains of science and  engineering during the last
two decades.  The desirable cryptographic properties of the chaotic maps such as
sensitivity  to  initial  conditions  and  random behavior  have  attracted  the
attention of researchers to develop  new PRNG with chaotic properties. Recently,
many  scholars have  made an  effort to  investigate chaotic  PRNGs in  order to
promote communication security~\cite{Behnia20113455}~\cite{Niansheng}~\cite{Sun20092216}.  
One of the existing chaos-based  PRNGs is using
chaotic iterations schemes.

A  short overview of  our recently  proposed PRNGs  based on  Chaotic Iterations
are      given      hereafter.       In
Ref.~\cite{Bahi2008},  it is proven  that chaotic  iterations (CIs),  a suitable
tool for fast computing  iterative algorithms, satisfies the topological chaotic
property, as it is defined by Devaney~\cite{Dev89}.  The chaotic behavior of CIs
is exploited in~\cite{bgw09:ip}, in order  to obtain an unpredictable PRNG that
depends on two  logistic maps.   The resulted  PRNG shows
better   statistical   properties   than   each  individual   component   alone.
Additionally, various chaos properties  have been established.  The advantage of
having  such chaotic  dynamics  for PRNGs  lies,  among other  things, in  their
unpredictability character.  These chaos properties, inherited from CIs, are not
possessed by  the two inputted generators.   We have shown that,  in addition of
being  chaotic,  this generator  can  pass the  NIST  battery  of tests,  widely
considered as a  comprehensive and stringent battery of  tests for cryptographic
applications~\cite{RSN+10}.              Then,             in            the
papers~\cite{guyeuxTaiwan10,bgw10:ip}, we have achieved  to improve the speed of
the former  PRNG by replacing  the two logistic  maps: we used two  XORshifts 
in~\cite{guyeuxTaiwan10},   and    ISAAC   with   XORshift    
in~\cite{bgw10:ip}.
Additionally, we  have shown that  the first generator  is able to  pass DieHARD
tests~\cite{diehard},  whereas   the  second   one  can   pass  
TestU01~\cite{testU01}. 

In spite of the fact that all these previous algorithms are parametrized with 
the embed PRNG, they all iterate  
the same function namely, 
the  vectorial boolean  negation later  denoted as $\neg$.  
It is then judicious to investigate whether other functions may replace 
the $\neg$ function in the above approach.
In the positive case, the user should combine its own
function and its own PRNGs to provide a new PRNG instance.
The approach developed along these lines solves this issue 
by providing a class of functions whose iterations are chaotic 
according to Devaney and such that resulting PRNG success statistical tests.

The rest of this paper is organized  in the following way. In the next 
section, some basic definitions  concerning CIs are  recalled.
Then,  our family of generators based  on   discrete  CIs  is   
presented  in  Section~\ref{Basic prng recalls} with some improvements. 
Next, Section~\ref{sec:instantiating} gives a characterization of 
functions whose iterations are chaotic. A practical note presents 
an algorithm allowing to generate some instances of such functions. 
These ones are then embedded in the algorithm presented in
Sect.~\ref{sec:modif} where we show why generator of 
Sect.~\ref{Basic prng recalls} is not convenient for them. 
In Section~\ref{sec:nist}, various
tests are  passed with a goal  to decide whether all chaotic functions
are convenient in a PRNG context.
 The
paper ends with a conclusion section where our contribution is summarized and intended future work is presented.

\section{Discrete Chaotic Iterations: recalls}\label{Basic recalls}
Let us denote by $\llbracket a ; b \rrbracket$ the interval of integers:
$\{a, a+1, \hdots, b\}$. A boolean system (BS) is a collection of  $n$ components.  Each 
component $i \in \llbracket 1; n \rrbracket$ 
takes its  value $x_i$ among  the domain  $\Bool=\{0,1\}$.  A
\emph{configuration}  of the  system at  discrete  time $t$  
(also called  at \emph{iteration} $t$) is the vector
$x^{t}=  (x_1^{t},  \ldots,   x_{n}^{t})  \in
\Bool^n$.

The dynamics of the system is  described according to a  function $f :
\Bool^n \rightarrow \Bool^n$ such that:
$f(x) = (f_1(x), \ldots, f_n(x))$.

Let be given a configuration $x$. In what follows the configuration 
$N(i,x) = (x_1,\ldots,\overline{x_i},\ldots,x_n)$ 
is obtained by switching the $i-$th component of 
$x$. Intuitively, $x$ and $N(i,x)$ are neighbors.
The discrete iterations of the $f$ function are represented  
by the so called graph of iterations.

\begin{definition}[Graph of iterations]
In the oriented \emph{graph of iterations},
vertices are configurations of $\Bool^n$ and
there is an arc labeled $i$ from $x$ to  $N(i,x)$
iff $f_i(x)$ is $N(i,x)$ (we consider 1-bit transitions).
\end{definition}

In the  sequel, the \emph{strategy}  $S=(S^{t})^{t \in \Nats}$  is the
sequence of the components that may be updated at time $t$,
$S^{t}$ denotes the $t-$th term of the strategy $S$.

Let  us  now introduce  two  important  notations.  $\Delta $  is  the
\emph{discrete        Boolean        metric},        defined        by
$\Delta(x,y)=0\Leftrightarrow  x=y$,  and   the  function  $F_{f}$  is
defined  for  any  given  application
  $f:\Bool^{n}  \to \Bool^{n}$ by
$$\begin{array}{ccl}
F_{f}:  \llbracket1;n\rrbracket\times \Bool^{n}
& \rightarrow  & 
\Bool^{n} \\  
(s,x) & \mapsto  & \left(
x_{j}.\Delta   (s,j)+f_j(x).   \overline{\Delta  (s,j)}\right)_{j\in
\llbracket1; n \rrbracket},\end{array}$$

\noindent where the point and the line above delta are multiplication and negation respectively. With such a notation, configurations are defined for times $t=0,1,2,\ldots$
by:
\begin{equation}\label{eq:sync}   \left\{\begin{array}{l}   x^{0}\in
\Bool^{n} \textrm{ and}\\
 x^{t+1}= F_{f}(S^t,x^{t})
\end{array} \right.
\end{equation}

Finally, iterations of (\ref{eq:sync}) can be described by
the following system

\begin{equation} \left\{
\begin{array}{l} X^{0} 
= ((S^t)^{t \in \Nats},x^0) \in 
\llbracket1;n\rrbracket^{\Nats}\times\Bool^{n}\\ 
  X^{k+1}=G_{f}(X^{k}),
\end{array} \right. \label{eq:Gf}
\end{equation}
such that
\begin{equation*}     
G_{f}\left((S^t)^{t \in \Nats},x\right)    
=\left(\sigma((S^t)^{t \in \Nats}),F_{f}(S^0,x)\right),
\end{equation*}
where $\sigma$ is the function that returns
the strategy  $(S^t)^{t \in \Nats}$ where the first term (\textit{i.e.}, $S^0$) 
has been removed.
In other words, at the $t^{th}$ iteration, only the $S^{t}-$th cell is
modified; the resulting strategy is the initial one where 
the first $t$ terms have been removed.

A previous work~\cite{Bahi2008} has shown a fine metric space 
such  that iterations of the map $G_f$ are chaotic in the sense
of Devaney~\cite{Dev89} when $f$ is the negation function $\neg$.
This   definition   consists    of   three   conditions:   topological
transitivity,  density   of  periodic  points,   and  sensitive  point
dependence  on initial  conditions.
Topological transitivity  is  established when, for any element, 
any neighborhood of its future evolution 
eventually overlap with any other given region.
On  the contrary, a dense set  of periodic points
is an  element of  regularity that a  chaotic dynamical system  has to
exhibit.  This regularity ``counteracts'' the effects of transitivity.
Finally,  a system  is sensitive to initial conditions
if future evolution of any point in its neighborhood are significantly
different. 
This result theoretically  implies the "quality" of the randomness.

The next section formalizes with chaotic iterations terms 
the PRNG algorithm presented in~\cite{bgw09:ip}.

\section{Chaos based PRNG}\label{Basic prng recalls}
This section aims at formalizing a PRNG algorithm
already presented in~\cite{bgw09:ip} and gives some 
improvements.

First of all, Let us intorduce \textit{XORshift}, generator. Xorshift is a category of pseudorandom number generators 
designed by George Marsaglia \cite{Marsaglia2003} that repeatedly uses the transform of exclusive or on a number with a bit shifted version 
of itself. A XORshift operation is defined as follows.
\begin{algorithm}
\KwIn{the internal state $z$ (a 32-bits word)}
\KwOut{$y$ (a 32-bits word)}
$z\leftarrow{z\oplus{(z\ll13)}}$\;
$z\leftarrow{z\oplus{(z\gg17)}}$\;
$z\leftarrow{z\oplus{(z\ll5)}}$\;
$y\leftarrow{z}$\;
return $y$\;
\medskip
\caption{An arbitrary round of XORshift algorithm}
\label{XORshift}
\end{algorithm}

Then the design procedure of this generator is summed
up in Algorithm~\ref{Chaotic iteration1}.

\begin{algorithm}
\KwIn{an initial state $x^0$ ($n$ bits)}
\KwOut{a state $x$ ($n$ bits)}
$x\leftarrow x^0$\;
$k\leftarrow{\textit{reallocate}(\textit{XORshift}() \mod (2^n-1))}$\;
$x \leftarrow \textit{iterate\_G}(\textit{neg},\textit{XORshift},k, x)$\;
return $x$\;
\caption{An arbitrary round of the (\textit{XORshift},\textit{XORshift}) generator}
\label{Chaotic iteration1}
\end{algorithm}

Let be given a seed as the internal state $x$. This algorithm outputs
a random configuration $x'$. 
It  is based on the \textit{XORshift}, generator which is called 
in two situations. 
The first one occurs while generating the parameter of the $\textit{reallocate}$
function that aims at  computing the number $k$ of time a function has to be iterated. 
The second one occurs as a parameter of 
\textit{iterate\_G}, which executes the iterations of $G$ 
as defined in~(\ref{eq:Gf}), with $f = neg, S = XORshift, x$ as initial state, and $k$ for the number of iterations. 


Firstly, let us focus on the $\textit{reallocate}$ function, which is defined by: 
$$
\textit{reallocate}(k) = \left\{
\begin{array}{llrcl}
0~~~ & \textrm{ if }  &0  &\le k  < & \displaystyle{{n \choose 0}} \\
1 &  \textrm{ if } & \displaystyle{{n \choose 0}}& \le k  < &
\displaystyle{\sum_{i=0}^1{n \choose i}}\\
\vdots&&&\vdots&\\
n & \textrm{ if } &  \displaystyle{\sum_{i=0}^{n-1}{n \choose i}}
 & \le k  \le & 2^n-1
\\
\end{array}
\right.
$$

Formally, the set $\llbracket 0, 2^n-1 \rrbracket$ is partionned into 
subsets 
$\llbracket \Sigma_{i=0}^j{n \choose 0}  , 
            \Sigma_{i=0}^{j+1}{n \choose i} \llbracket$ 
where $j\in \llbracket 0, n-1 \rrbracket$.
Each interval bound is a binomial coefficient:
it gives the number of combinations of $n$ things taken $j$.
In our context, 
it is the number of configurations  
$(x_1,\ldots,x_n)$ that can be built by negating $j$ elements 
among $n$.
The function $\textit{reallocate}$ allows to compute a distribution on
$\llbracket 0, n \rrbracket$ 
that permits to reach configurations in $\llbracket 0, 2^n-1 \rrbracket$
uniformly.

Let us present now the $\textit{iterate\_G}$ function. 
It starts with computing the strategy $S$ of lenght $k$ as the result of 
a usual \textit{sample} (not detailled here) 
function that selects $k$ elements among $n$
following a PRNG $r$ given as the first parameter.  
The loop next reproduces $k$ iterations of $G_f$ as define in Equ.~(\ref{eq:Gf})

\begin{algorithm}
\KwIn{a function $f$, a PRNG $r$, 
  an iterations number $k$, a binary number $x^0$ ($n$ bits)}
\KwOut{a binary number $x$ ($n$ bits)}
$x\leftarrow x^0$\;
S = $\textit{sample}(r,k,n)$\;
\For{$i=0,\dots,k-1$}
{
  $s \leftarrow S[i]$\;
  $x\leftarrow  F_f(s,x) $\;
}
return $x$\;
\caption{The \textit{iterate\_G} function. \label{algo:it}}
\end{algorithm}

Compared to work~\cite{bgw09:ip}, this algorithm is:
\begin{itemize}
\item close to the formal iterations of $G_f$: strategy is explicitely 
  computed and there are as many iterations as the number of executed loops. 
\item more efficient: in the previous work, loops are 
  executed untill $k$ distinct elements have been switched leading to possibly
  more iterations. In the opposite, the function \textit{iterate\_G} exactly
  executes $k$ loops when $k$ iterations are awaited. However, this improvement
  moves the problem into the \textit{sample} function, which is classically
  tuned to speed up its global behavior. In such a context we take a benefit 
  of this improvement. Table~\ref{tabl:eff} compares these two algorithms 
  in terms of execution time with respect to the number of generated elements.
  The improvement is about 9\%.
\end{itemize}

\begin{table}
\centering
\begin{tabular}{|r|r|r|r|r|r|}
  \hline
  & 100 & 10000 & 100000 & 1000000 & 1000000\\
  \hline
  Speed up & 
10\% & 7.8 \%&  8.8 \%& 8.1\% &  9.5\%  \\
  \hline
\end{tabular}
\caption{Speed up improvement from Algorithm~\cite{bgw09:ip}\label{tabl:eff}}
\end{table}

However as noticed in introduction, the whole (theoretical and practical) 
approach  is based on the negation function. The following section 
studies whether other functions can theoretically replace this one.

\section{Characterizing and Computing Functions for PRNG}\label{sec:instantiating}
This section presents other functions that theoretically could replace the 
negation function $\neg$ in the previous algorithms. 

In this algorithm and from the graph point of view, 
iterating the function $G_f$ from a configuration $x^0$
and according to a strategy $(S^t)^{t \in \Nats}$  
consists in traversing the directed iteration graph $\Gamma(f)$
from a vertex $x^0$ following the edge labelled with $S^0$, $S^1$, \ldots
Obviously, if some vertices cannot be reached from other ones,
their labels expressed as numbers cannot be output by the generator.
The \emph{Strongly connected component of $\Gamma(f)$}
(\textit{i.e.}, when there is a path from each vertex to every other one),
denoted by SCC in the following~\cite{ita09},
is then a necessary condition for the function $f$.   
The following result shows this condition is sufficient to make 
iterations of $G_f$ chaotic.

\begin{theorem}[Theorem III.6, p. 91 in~\cite{GuyeuxThese10}]
\label{Th:Caractérisation   des   IC   chaotiques}  
Let  $f$ be a function from $\Bool^{n}$ to  $\Bool^{n}$. Then 
$G_f$ is chaotic  according to  Devaney iff the graph
$\Gamma(f)$ is strongly connected.
\end{theorem}

Any function such that the graph $\Gamma(f)$ is strongly connected
is then a candidate for being iterated in $G_f$
for pseudo random number generating. 
Thus, let us show how to compute a map $f$ 
with a strongly connected graph of iterations $\Gamma(f)$.

We first consider the negation function $\neg$. The iteration graph 
$\Gamma(\neg)$ is obviously strongly connected:
since each configuration $(x_1,\ldots, x_n)$ may reach one of its $n$ neighbors,
there is then a bit by bit path from any 
$(x_1,\ldots, x_n)$ to any $(x'_1,\ldots, x'_n)$.
Let then $\Gamma$ be a graph, initialized with $\Gamma(\neg)$, 
the algorithm iteratively does the two following stages: 
\begin{enumerate}
\item select randomly an edge of the current iteration graph $\Gamma$ and
\item check whether the current iteration graph without that edge 
  remains strongly connected (by a Tarjan algorithm~\cite{Tarjanscc72}, for instance). In the positive case the edge is removed from $G$,
\end{enumerate}
until a rate $r$ of removed edges is greater
than a threshold given by the user.

Formally, if $r$ is close to $0\%$ (\textit{i.e.}, few edges are removed), 
there should remain about $n\times 2^n$ edges (let us recall that $2^n$ is the amount of nodes).
In the opposite case, if $r$ is close to $100\%$, there are left about $2^n$
edges.
In all the cases, this step returns the last graph $\Gamma$ 
that is strongly connected.  
It is not then obvious to return the function $f$ whose iteration
graph is $\Gamma$.

However, such an approach suffers from generating many functions with similar
behavior due to the similarity of their graph.    More formally, let us recall
the graph isomorphism definition that resolves this issue. 
Two directed graphs $\Gamma_1$ and $\Gamma_2$
 are \emph{isomorphic} 
if there exists a permutation $p$ from the vertices of
$\Gamma_1$ to the vertices of $\Gamma_2$ such that
there is an arc from vertex $u$ to vertex  $v$  in $\Gamma_1$ 
iff there is an arc from vertex $p(u)$ to vertex  $p(v)$  in 
$\Gamma_2$.

Then, let $f$ be a function, $\Gamma(f)$ 
be its iteration graph, and $p$ 
be a permutation of vertices of $\Gamma(f)$. 
Since $p(\Gamma(f))$ and $\Gamma(f)$ are isomorphic,   
then iterating  $f$ 
(\textit{i.e.}, traversing $\Gamma(f)$) from the initial configuration $c$
amounts to iterating the function whose iteration graph is $p(\Gamma(f))$
from the configuration $p(c)$.
Graph isomorphism being an equivalence relation, the sequel only 
consider the quotient set of functions with this relation over their graph.
In other words, two functions are distinct if and only if their iteration
graph are not isomorphic.

\begin{table}
\centering
\begin{tabular}{|c|c|c|}
\hline
Function $f$ & $f(x)$, for $x$ in $(0,1,2,\hdots,15)$ & Rate\\ 
\hline
$\neg$&(15,14,13,12,11,10,9,8,7,6,5,4,3,2,1,0)&0\%\\
\hline
$\textcircled{a}$&(15,14,13,12,11,10,9,8,7,6,7,4,3,2,1,0)&2.1\%\\
\hline
$\textcircled{b}$&(14,15,13,12,11,10,9,8,7,6,5,4,3,2,1,0)&4.1\%\\
\hline
$\textcircled{c}$&(15,14,13,12,11,10,9,8,7,7,5,12,3,0,1,0)&6.25\%\\
\hline
$\textcircled{d}$&(14,15,13,12,9,10,11,0,7,2,5,4,3,6,1,8)&16.7\%\\
\hline
$\textcircled{e}$&(11,2,13,12,11,14,9,8,7,14,5,4,1,2,1,9)&16.7\%\\
\hline
$\textcircled{f}$&(13,10,15,12,3,14,9,8,6,7,4,5,11,2,1,0)&20.9\%\\
\hline
$\textcircled{g}$&(13,7,13,10,11,10,1,10,7,14,4,4,2,2,1,0)&20.9\%\\
\hline
$\textcircled{h}$&(7,12,14,12,11,4,1,13,4,4,15,6,8,3,15,2)&50\%\\
\hline
$\textcircled{i}$&(12,0,6,4,14,15,7,15,11,1,14,2,7,4,7,9)&75\%\\
\hline
\end{tabular}
\caption{Functions with SCC graph of iterations\label{table:nc}}
\end{table}

Table~\ref{table:nc} presents generated functions 
that have been ordered by the rate of removed edges in their
graph of iterations compared to the iteration graph $\Gamma(\neg)$
of the boolean negation function $\neg$.

For instance let us consider  the function $\textcircled{g}$ from $\Bool^4$ to $\Bool^4$
defined by the following images: 
$[13,7,13,10,11,10,1,10,7,14,4,4,2,2,1,0]$.
In other words,  the image of $3 ~(0011)$ by $\textcircled{g}$ is $10 ~(1010)$: it is obtained
as  the  binary  value  of  the  fourth element  in  the  second  list
(namely~10).  It  is not  hard to verify  that $\Gamma(\textcircled{d})$ is  SCC.
Next section gives practical evaluations of these functions.

\section{Modifying the PRNG Algorithm}\label{sec:modif}
A coarse attempt could directly embed each function of table~\ref{table:nc}  
in the $\textit{iterate\_G}$ function defined in Algorithm~\ref{algo:it}.
Let us show the drawbacks of this approach on a more simpler example.

\begin{figure}[ht]
  \centering
  \subfloat[Negation]{
    \includegraphics[width=1.33cm]{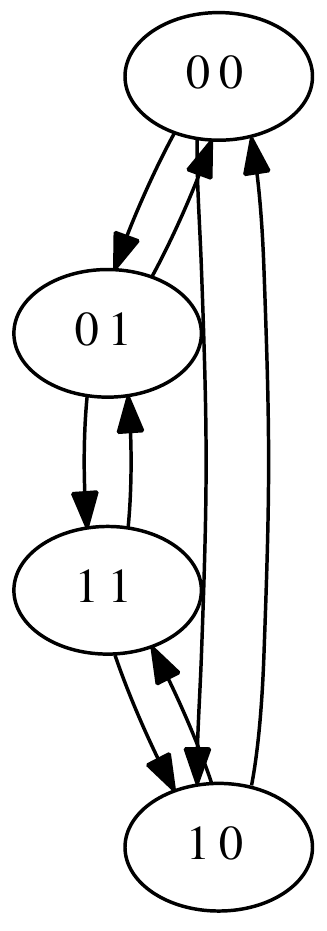}
    \label{fig:comp:n}
  }
  \subfloat[$(1,3,0,2)$]{
    \includegraphics[width=1.62cm]{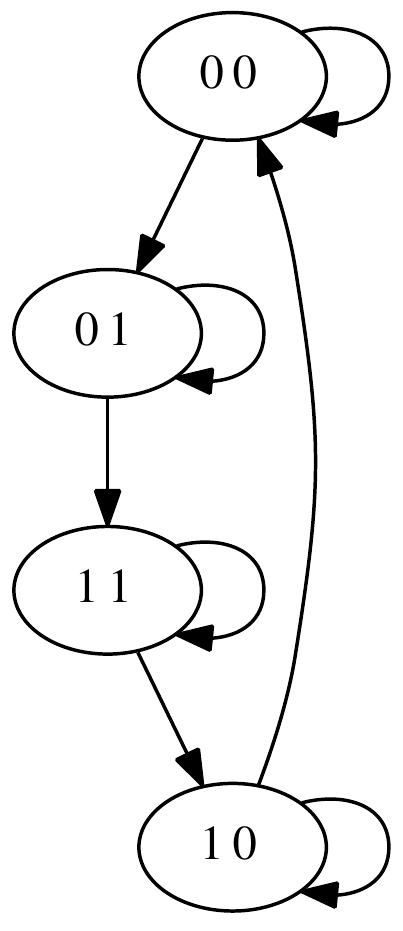}
    \label{fig:comp:f}
  }
  \caption{Graphs of Iterations}
    \label{fig:xplgraph}
\end{figure}

Let us consider for instance $n$ is two, the negation function on $\Bool^2$, and
the function $f$ defined by the list $[1,3,0,2]$ (i.e., $f(0,0) = (0,1), f(0,1) = (1,1), f(1,0) = (0,0),$ and $f(1,1)=(1,0)$) whose iterations graphs are represented 
in Fig.~\ref{fig:xplgraph}.
The two graphs are strongly connected and thus the vectorial negation function 
should theoretically  be replaced by the function $f$.

In the graph of iterations $\Gamma({\neg})$ (Fig.~\ref{fig:comp:n}), 
let us compute the probability $P^t_{\neg}(X)$ to reach the node $X$ in $t$ iterations 
from the node 00. Let $X_0$, $X_1$, $X_2$, $X_3$ be the nodes   
$00$, $01$, $10$ and $11$.
For $i\in \llbracket 0,3 \rrbracket$,  $P^1_{\neg}(X_i)$,  are respectively equal to 
0.0, 0.5, 0.0, 0.5. In two iterations   $P^2_{\neg}(X_i)$
are 0.5, 0.0, 0.5, 0.0.
It is obvious to establish that we have 
$P^{2t}(X_i) = P^{0}(X_i)$ and $P^{2t+1}(X_i) = P^{1}(X_i)$ for any $t\in \Nats$.
Then in $k$ or $k+1$ iterations all these probabilities are equal to 0.25.  

Let us apply a similar reasoning for the function $f$ defined by $[1,3,0,2]$.
In its iterations graph $\Gamma(f)$ (Fig.~\ref{fig:comp:f}),
and with $X_i$ defined as above,
the probabilities $P^1_{f}(X_i)$ to reach the node $X_i$ 
in one iteration  from the node 00
are respectively equal to 
0.5, 0.5, 0.0, 0.0.  
Next, probabilities  $P^2_{f}(X)$  are 0.25, 0.5, 0.25, 0.0. 
Next, $P^3_{f}(X)$  are 0.125, 0.375, 0.375, 0.125.
For each iteration, we compute the average deviation rate $R^t$ 
with 0.25 as follows.
$$
R^t= \dfrac{ \Sigma_{i=0}^3 \mid P^t_{f}(X_i)-0.25 \mid}
{4}.
$$  
The higher is this rate, the less the generator may uniformly reach any $X_i$ from $00$.
For this example, it is necessary to iterate 14 times in order to
observe a deviation from 0.25 less 
than 1\%. 
A similar reasoning has been applied for all the functions listed in Table~\ref{table:nc}.
The table~\ref{tab:dev} summarizes their deviations with uniform distribution and gives the 
smallest iterations number the smallest deviation has been obtained. 

\begin{table}
\centering
\begin{tabular}{|c|r|r|}
\hline
Name & Deviation & Suff. number of it. \\

\hline
$\textcircled{a}$ &  8.1\% & 167 \\
\hline
$\textcircled{b}$ &  1\%  & 105 \\
\hline
$\textcircled{c}$ &  18\% & 58 \\
\hline
$\textcircled{d}$ &  1\% & 22  \\
\hline
$\textcircled{e}$ &  24\% & 19 \\
\hline
$\textcircled{f}$ &  1\%  & 14 \\
\hline
$\textcircled{g}$ &  20\% &  6 \\
\hline
$\textcircled{h}$ & 45.3\% & 7 \\
\hline
$\textcircled{i}$ & 53.2\%& 14 \\
\hline
\end{tabular}
\caption{Deviation with Uniform Distribution \label{tab:dev}}
\end{table}

With that material we present in Algorithm~\ref{CI Algorithm}
the method that allows to take any chaotic function as 
the core of a pseudo random number generator.
Among the parameters, it takes the number $b$ of minimal iterations
that have to be executed to get a uniform like 
distribution. For our experiments $b$ is set with the value 
given in the third column of Table~\ref{tab:dev}.

\begin{algorithm}
\KwIn{a function $f$, an iteration number $b$, an initial state $x^0$ ($n$ bits)}
\KwOut{a state $x$ ($n$ bits)}
$x\leftarrow x^0$\;
$k\leftarrow b + (\textit{XORshift}() \mod 2)$\;
\For{$i=0,\dots,k-1$}
{
$s\leftarrow{\textit{XORshift}() \mod n}$\;
$x\leftarrow{F_f(s,x)}$\;
}
return $x$\;
\medskip
\caption{modified PRNG with various functions}
\label{CI Algorithm}
\end{algorithm}

Compared to the algorithm~\ref{Chaotic iteration1}
parameters of this one are 
the function $f$ to embed and 
the smallest number of time steps $G_f$ is iterated. 
First, the number of iterations is either $b$ or $b+1$ depending on the 
value of the \textit{XORshift} output (if the next value .
Next, a loop that iterates $G_f$ is executed.

In this example, $n$ and $b$ are equal to $4$ for easy understanding.
The initial state of the system $x^0$ can be seeded by the decimal part of the current time.
For example, the current time in seconds since the Epoch is 1237632934.484088,
so $t = 484088$. $x^0 = t \mod 16 $ in binary digits, then $x^0 = 0100$.
$m$ and $S$ can now be computed from \textit{XORshift}.
\begin{itemize}
\item $f$ = [14,15,13,12,11,10,9,8,7,6,5,4,3,2,1,0] 
\item $k$ = 4, 5, 4,\ldots
\item $s$ = 2,  4,  2,  3, ,  4,  1,  1,  4,  2, ,  0,  2,  3,  1,\ldots
\end{itemize}
Chaotic iterations are done with initial state $x^0$,
the mapping function $f$, and strategy $s^1$, $s^2$\ldots
The result is presented in Table \ref{table application example}. 
Let us recall that sequence $k$ gives the states $x^t$ to return: $x^4, x^{4+5}, x^{4+5+4}$\ldots
Successive stages are detailed in Table~\ref{table application example}.

\begin{table*}[t]
\centering
\begin{tabular}{|c|ccccc|cccccc|ccccc|}
\hline\hline
$k$ &  \multicolumn{5}{|c|}{4} &  \multicolumn{6}{|c|}{5} & \multicolumn{5}{|c|}{4}\\
\hline
$s$ & 2 & 4 & 2 & 3 & & 4 & 1 & 1 & 4 & 2 & & 0 & 2 & 3 & 1 &  \\ \hline
&$f(4)$&$f(0)$&$f(0)$&$f(4)$ &  &$f(6)$ &$f(7)$ &$f(15)$ &$f(7)$ &$f(7)$ & 
&$f(2)$ &$f(0)$ & $f(4)$& $f(6)$&  \\
\multirow{4}{*}{$f$} 
 & 1& 1& 1& 1&
 & 1& \textbf{1} & \textbf{0} &1 &1 &
 &1 & 1& 1&\textbf{1} &
  \\
 & \textbf{0} & 1& \textbf{1} & 0 &
 &0 &0& 0&0 & \textbf{0}&
 &1 &\textbf{1} & 0&0 & \\
 &1 & 1& 1& \textbf{1}&
 &0 &0 &0 &0 &0 &
& \textbf{0} & 1& \textbf{1} & 0 &
  \\
 &1 &\textbf{0} &0 &1 &
 &\textbf{1} &0 &0 &\textbf{0} &0 &
 &1 &0 & 1&1 &
 \\\hline
$x^{0}$ & & & & & $x^{4}$ & & & & & & $x^{9}$ & & & & & $x^{13}$  \\
4 &0 &0 &4 &6&6 &7 &15 &7 &7 &7 &2&0  &4 &6 &14 &14   \\ 
& & &  & & && & & & & & &  & & &   \\
0 & & & & &
0 & & $\xrightarrow{1} 1$ & $\xrightarrow{1} 0$ & & &
0 & & & & $\xrightarrow{1} 1$ &
1  \\
1 & $\xrightarrow{2} 0$ & & $\xrightarrow{2} 1$ &  &
1 & & & & & $\xrightarrow{2} 0$ &
0 & & $\xrightarrow{2} 1$ & & & 
1 \\
0 & & & & $\xrightarrow{3} 1$ &
1 & & & & & &
1 & $\xrightarrow{3} 0$ & & $\xrightarrow{3} 1$ &  &
1 \\
0 & & $\xrightarrow{4} 0$ & & &
0 &$\xrightarrow{4} 1$ & & & $\xrightarrow{4} 0$& &
0 & & & & &
0 \\
\hline\hline
\end{tabular}
\caption{Application example}
\label{table application example}
\end{table*}

\begin{figure}
\centering
 \subfloat[Function $\textcircled{e}$]{
   \includegraphics[width=7cm]{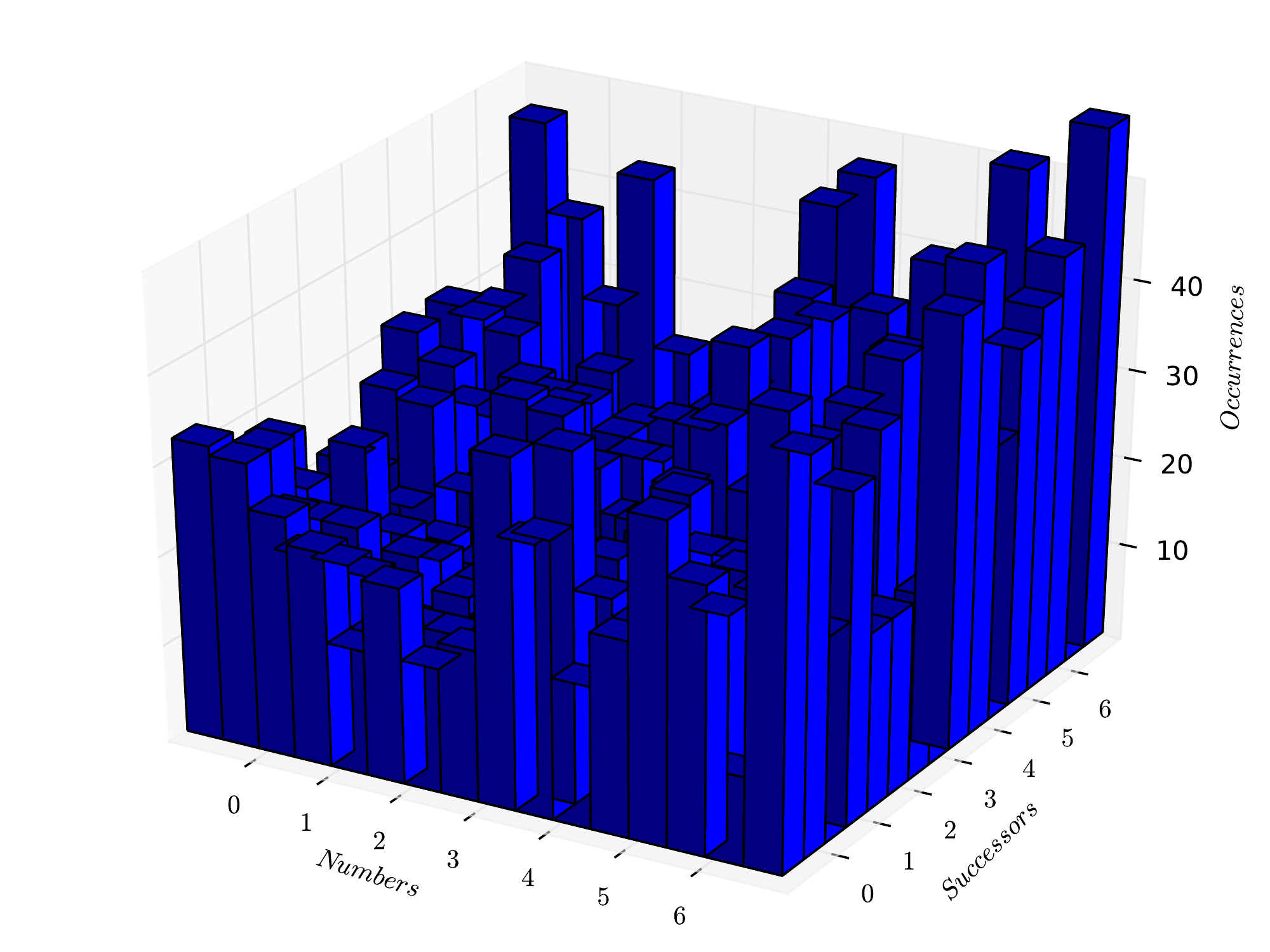}
   \label{fig:f5}
 }
$\qquad$
\subfloat[Function $\textcircled{f}$]{
  \includegraphics[width=7cm]{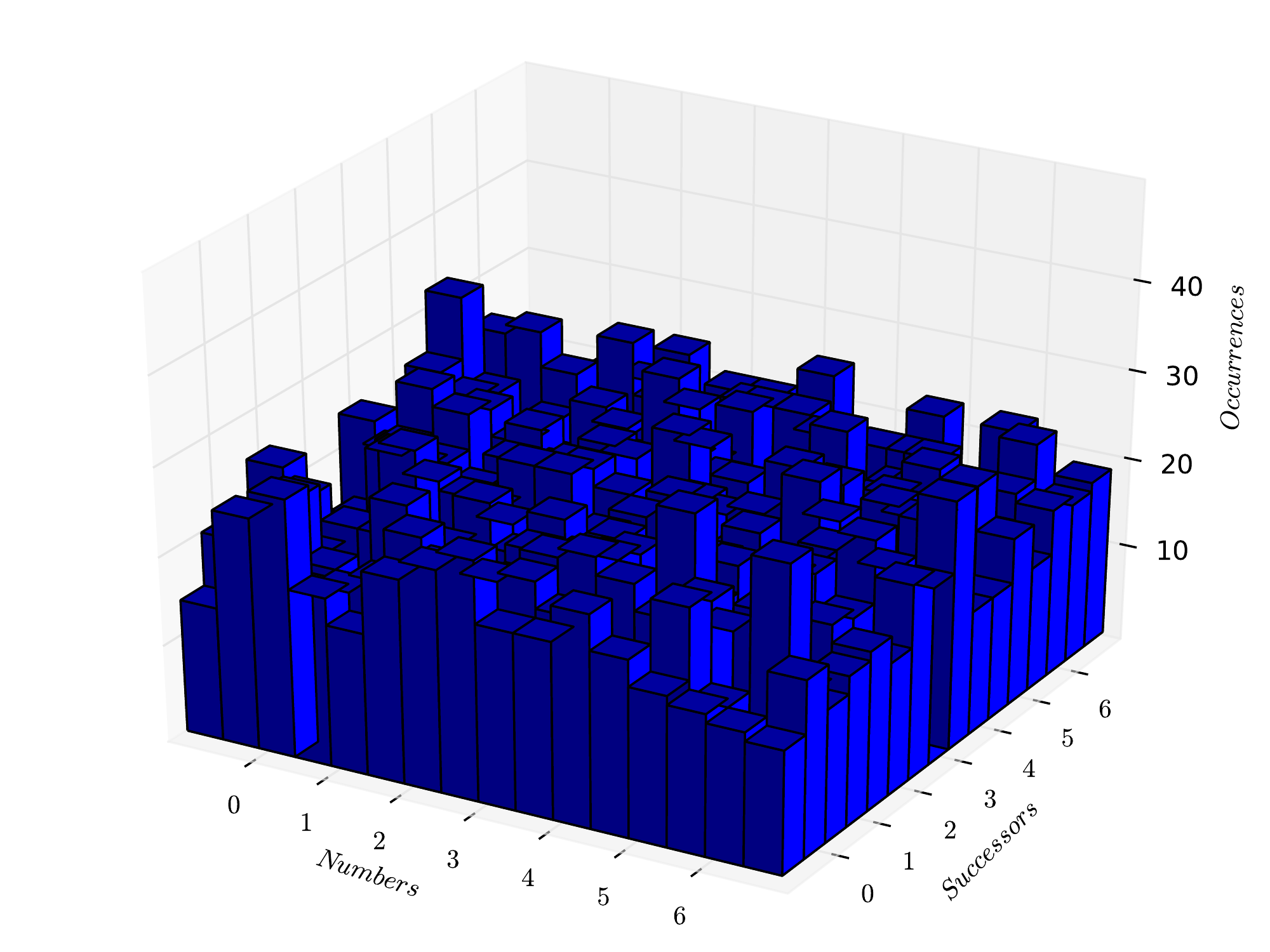}
     \label{fig:f6}

}
\caption{Repartition of function outputs.} \label{fig:fs}
\end{figure}

To illustrate the deviation, Figures~\ref{fig:f5} and~\ref{fig:f6} represent
the simulation outputs of 5120 executions with  $b$ equal to $40$
for $\textcircled{e}$ and $\textcircled{f}$ respectively.
In these two figures, the point $(x,y,z)$ can be understood as follows.
$z$ is the number of times the value $x$ has been succedded by the value $y$ in the 
considered generator.
These two figures explicitly confirm that outputs of functions $\textcircled{f}$ are 
more uniform that these of the function $\textcircled{e}$. 
In the former each number $x$ reaches about 20 times each number $y$ whereas
in the latter, results vary from 10 to more that 50.

\section{Experiments}\label{sec:nist}

A convincing  way to prove the  quality of the produced
sequences  is to confront them  with the NIST
(National  Institute of  Standards  and Technology)  Statistical  Test Suite  SP
800-22~\cite{RSN+10}.
This  is  a  statistical  package consisting  of  15 tests that
focus on  a variety of different types of non-randomness
that could occur in a (arbitrarily  long) binary  sequences
produced  by a pseudo-random  number  generators.


For all 15 tests, the significance level $\alpha$ was set to $1\%$. If a p-value is greater than 0.01, the keystream is accepted as random with a confidence of $99\%$; otherwise, it is considered as non-random.
For each statistical test, a set of p-values is produced from a set of sequences obtained by our generator (i.e., 100 sequences are generated and tested, hence 100 p-values are produced). 

Empirical results can be interpreted in various ways. In this paper, 
we check whether $\mathbb{P}_T$ (P-values of p-values), which arise via the application of a chi-square test, were all higher than 0.0001. This means
that all p-values are uniformly distributed over (0, 1) interval as expected for an ideal random number generator.

\begin{table*}[t]
\renewcommand{\arraystretch}{1.3}
\caption{NIST SP 800-22 test results ($\mathbb{P}_T$)}
\label{The passing rate}
\centering
  \begin{tabular}{|l||c|c|c|c|c|c|c|c|c|}
    \hline
Method &$\textcircled{a}$& $\textcircled{b}$ &  $\textcircled{c}$ & $\textcircled{d}$ & $\textcircled{e}$ & $\textcircled{f}$ & $\textcircled{g}$ & $\textcircled{h}$ & $\textcircled{i}$\\ \hline\hline

Frequency (Monobit) Test			&0.00000 &  0.45593 &  0.00000 &  0.38382 &  0.00000 &  0.61630 &  0.00000 &  0.00000 &    0.00000 \\ \hline
Frequency Test within a Block 			&0.00000 &  0.55442 &  0.00000 &  0.03517 &  0.00000 &  0.73991 &  0.00000 &  0.00000 &    0.00000 \\ \hline
Cumulative Sums (Cusum) Test* 			&0.00000 &  0.56521 &  0.00000 &  0.19992 &  0.00000 &  0.70923 &  0.00000 &  0.00000 &    0.00000 \\ \hline
Runs Test					&0.00000 &  0.59554 &  0.00000 &  0.14532 &  0.00000 &  0.24928 &  0.00000 &  0.00000 &    0.00000 \\ \hline				
Test for the Longest Run of Ones in a Block	&0.20226 &  0.17186 &  0.00000 &  0.38382 &  0.00000 &  0.40119 &  0.00000 &  0.00000 &    0.00000 \\ \hline
Binary Matrix Rank Test				&0.63711 &  0.69931 &  0.05194 &  0.16260 &  0.79813 &  0.03292 &  0.85138 &  0.12962 &    0.07571 \\ \hline
Discrete Fourier Transform (Spectral) Test 	&0.00009 &  0.09657 &  0.00000 &  0.93571 &  0.00000 &  0.93571 &  0.00000 &  0.00000 &    0.00000 \\ \hline
Non-overlapping Template Matching Test*		&0.12009 &  0.52365 &  0.05426 &  0.50382 &  0.02628 &  0.50326 &  0.06479 &  0.00854 &    0.00927 \\ \hline
Overlapping Template Matching Test		&0.00000 &  0.73991 &  0.00000 &  0.55442 &  0.00000 &  0.45593 &  0.00000 &  0.00000 &    0.00000 \\ \hline
Maurer’s “Universal Statistical” Test		&0.00000 &  0.71974 &  0.00000 &  0.77918 &  0.00000 &  0.47498 &  0.00000 &  0.00000 &    0.00000 \\ \hline
Approximate Entropy Test			&0.00000 &  0.10252 &  0.00000 &  0.28966 &  0.00000 &  0.14532 &  0.00000 &  0.00000 &    0.00000\\ \hline
Random Excursions Test*				&NaN &  0.58707 &NaN &  0.41184 &NaN &  0.25174 &NaN &NaN &NaN \\ \hline
Random Excursions Variant Test*			&NaN &  0.32978 &NaN &  0.57832 &NaN &  0.31028 &NaN &NaN &NaN \\ \hline
Serial Test* (m=10)				&0.11840 &  0.95107 &  0.01347 &  0.57271 &  0.00000 &  0.82837 &  0.00000 &  0.00000 &    0.00000 \\ \hline
Linear Complexity Test				& 0.91141 &  0.43727 &  0.59554 &  0.43727 &  0.55442 &  0.43727 &  0.59554 &  0.69931 &    0.08558 \\ \hline
Success 					&5/15&15/15&4/15&15/15&3/15&15/15&3/15&3/15&3/15  \\ \hline
Computational time				&66.0507&47.0466&32.6808&21.6940&20.5759&19.2052&16.4945&16.8846&19.0256\\ \hline
  \end{tabular}
\end{table*}

Table~\ref{The  passing rate}  shows $\mathbb{P}_T$  of the  sequences  based on
discrete chaotic  iterations using  different ``iteration'' functions.  If there
are  at least  two statistical  values in  a test,  the test  is marked  with an
asterisk  and the  average value  is  computed to  characterize the  statistical
values. Here,  NaN means  a warning that  test is  not applicable because  of an
insufficient  number of cycles.  Time (in  seconds) is  related to  the duration
needed by each  algorithm to generate a $10^8$ bits long  sequence. The test has
been conducted using  the same computer and compiler  with the same optimization
settings for both algorithms, in order to make the test as fair as possible.

Firstly, the computational  time in  seconds has increased due to the
growth of the sufficient iteration numbers, as precised  in Table~\ref{tab:dev}.
For  instance, the fastest generator is $\textcircled{g}$ since each new 
number generation only requires 6 iterations.
Next, concerning the  NIST tests results, 
best situations are given by $\textcircled{b}$,  $\textcircled{d}$ and
$\textcircled{f}$. In the opposite, it can be observed that among the 15 tests,
less than 5 ones are a successful for  other functions. 
Thus,  we can draw a  conclusion that, $\textcircled{b}$, $\textcircled{d}$,
and $\textcircled{f}$ are qualified to be good PRNGs with chaotic property.
NIST tests results are not a surprise:
$\textcircled{b}$,  $\textcircled{d}$, and $\textcircled{f}$ have indeed a deviation less than 1\% with 
the uniform distribution as already precised in Table~\ref{tab:dev}.
The rate of removed edge in the graph $\Gamma(\neg)$ is then not a pertinent
criteria compared to the deviation with the uniform distribution property:
the function $\textcircled{a}$ whose graph $\Gamma(\textcircled{a})$ is $\Gamma(\neg)$ without the
edge $1010 \rightarrow 1000$ (\textit{i.e.}, with only one edge less than
$\Gamma(\neg)$) has dramatic results compared to the function 
$\textcircled{f}$ with many edges less.

Let us then try to give a characterization of convenient function. 
Thanks to a comparison with the other functions, we notice that 
$\textcircled{b}$,  $\textcircled{d}$, and $\textcircled{f}$ are composed of all the elements of
$\llbracket 0;15  \rrbracket$.
It means that $\textcircled{b}$,  $\textcircled{d}$, and $\textcircled{f}$, and even the vectorial  boolean
negation function are arrangements  of 
$\llbracket  0;2^n \rrbracket$  ($n=4$ in  this article)  into a
particular order.

\section{Conclusion}\label{sec:concl}
In this work we first have formalized the PRNG already presented in a previous 
work. It results a new presentation that has allowed to
optimize some part and thus has led to a more efficient algorithm.
But more fundamentally, this PRNG closely follows  iterations
that have been proven to be topological chaotic. 

By considering a characterization of functions with topological chaotic
behavior (namely those with a strongly connected graph of iterations),
we have  computed a new class of PRNG based on instances of such 
functions. 
These functions have been randomly generated starting from the negation
function. 
Then an  a posteriori analysis has checked whether any number may be 
equiprobabilistically reached from any other one. 

The NIST statistical test has confirmed that functions without 
equiprobabilistical behavior are not good candidates for
being iterated in our PRNG.
In the opposite, the other ones have topological chaos property
and success all the NIST tests.
To summarize the approach, all our previous approaches were based on 
only one function (namely the negation function) whereas we provide now 
a class of many trustworthy PRNG.

Future work are mainly twofold.
We will firstly study sufficient conditions 
to obtain functions with the two  properties of
equiprobability and strongly connectivity of its graph of iterations.
With such a condition any user should choose its own trustworthy PRNG.
Dually, we will continue the evaluation of randomness quality
by checking other  statistical series like 
DieHard\cite{diehard}, TestU01~\cite{testU01}\ldots on newly 
generated functions.

\enlargethispage{-50mm}

\bibliographystyle{plain}
\bibliography{mapasbase,abbrev,biblioand}

\end{document}